\documentclass[conference]{IEEEtran}%
\pdfoutput=1
\usepackage{amsfonts}
\usepackage{amsmath}
\usepackage{amssymb}
\usepackage{graphicx}%
\providecommand{\U}[1]{\protect\rule{.1in}{.1in}}
\IEEEoverridecommandlockouts
\newtheorem{theorem}{Theorem}

\begin{document}

\title{Performance of polar codes for quantum and private classical communication}
\author{\IEEEauthorblockN{Zachary Dutton and Saikat Guha} \IEEEauthorblockA{\it Raytheon BBN Technologies,\\Cambridge, Massachusetts, USA 02138} \and \IEEEauthorblockN{Mark M.\ Wilde} \IEEEauthorblockA{\it School of Computer Science, McGill University\\Montreal, Quebec, Canada} }

\maketitle

\begin{abstract}
We analyze the practical performance of quantum polar codes, by 
computing rigorous bounds on block error probability 
and by numerically simulating them. We evaluate our bounds for quantum erasure channels with
coding block lengths between $2^{10}$ and $2^{20}$,
and we report the results of simulations for quantum erasure channels, quantum
depolarizing channels, and ``BB84'' channels with coding block lengths up to $N=1024$. For quantum
erasure channels, we observe that high quantum data rates
can be achieved for block error rates $P_{e} \leq 10^{-4}$ and that somewhat lower quantum data
rates can be achieved for quantum depolarizing and BB84 channels. Our results here also
serve as bounds for and simulations of private classical data transmission over these
channels, essentially due to Renes' duality bounds for privacy amplification
and classical data transmission of complementary observables. Future work
might be able to improve upon our numerical results for quantum depolarizing
and BB84 channels by employing a polar coding rule other than the heuristic used here.

\end{abstract}


Arikan's polar codes are an important recent development in coding
theory \cite{A09}. These codes provably achieve the Shannon capacity of symmetric
binary-input memoryless channels and
have an $O(N \log N)$ complexity for both encoding and decoding,
where $N$ is the number of channel uses. Polar codes exploit the channel
polarization effect, in which a particular recursive encoding induces a set of
virtual channels, such that some of the virtual channels are near perfect for
data transmission and the others are near useless for this task. The fraction
of the virtual channels that are near perfect is equal to the symmetric
capacity of the original channel.

Several authors have now extended the ideas of Arikan to the
domain of quantum information theory, in order to accomplish a variety of
information processing tasks, including classical data transmission
\cite{WG11,GW12}, private classical data transmission \cite{WG11a,WR12a}, and
quantum data transmission \cite{RDR11,WG11a,WR12}. Among these works,
Ref.~\cite{RDR11} demonstrated that all of the above impressive features of
classical polar codes are preserved when constructing quantum polar codes for
sending quantum data over quantum Pauli channels or quantum erasure channels.
Refs.~\cite{WG11a,WR12a} then followed up with similar results for private
classical data transmission over these channels. The importance of Pauli
channels stems from the fact that any noisy qubit channel can be ``twirled'' to a
Pauli channel with a quantum capacity not higher than the original channel. A
quantum depolarizing channel is a Pauli channel that is often used as a
``worst-case scenario'' noise model, and a quantum erasure channel is a
simplified model of photon loss.

In light of the above theoretical works, it seems natural now to assess the
practical performance of finite-blocklength quantum polar codes on any of
the aforementioned channels. One could get a better idea of this by deriving
rigorous bounds on performance like Arikan's bounds for classical erasure
channels \cite{A09}, and another way to do so is simply by simulating their
performance numerically. Indeed, since the quantum decoder for Pauli and
erasure channels is a coherent version of Arikan's successive cancellation
decoder combined with other efficient operations \cite{RDR11}, it is possible
to have an efficient numerical simulation of these codes' performance.

In this paper, we derive bounds on the performance of quantum polar codes
on quantum erasure channels, and we report the results of numerical simulations determining the
performance of quantum polar codes on quantum erasure, depolarizing, and BB84
channels. In our numerical simulations, we consider the
performance of quantum successive cancellation decoding for block
lengths of 64, 256, and 1024. We find that with these moderate block lengths,
relatively large rates can be achieved on the quantum erasure
channel with a block error rate $P_{e} \leq 10^{-4}$. For the depolarizing
and BB84 channels, we compare our results against the well-known hashing bound for
quantum data transmission. For these channels, we discuss how our results serve
equally as simulations of polar codes for private classical data transmission.

We structure this paper as follows. To start, Sections
\ref{sec:classical-review}, \ref{sec:polar-review-quantum} and
\ref{sec:polar-review-private} review some of the recent advances in polar
coding for quantum channels \cite{WG11,RDR11,WG11a,WR12,WR12a,GW12}, focusing
on the approaches given in Refs.~\cite{WG11,WR12,WR12a}. Section~\ref{sec:erasure-bounds}
derives our bounds on quantum polar code performance for quantum erasure channels.
We then detail the results of our
numerical simulations in Section~\ref{sec:numerics}. In
Section~\ref{sec:private-app}, we discuss how our results apply in the setting
of private classical communication, and we finally conclude in
Section~\ref{sec:conclusion}.

\section{Polar codes for classical communication}

\label{sec:classical-review} Consider a channel $W$ with a classical input
$x\in\left\{  0,1\right\}  $ and a quantum output $\rho_{x}$:%
\begin{equation}
W:x\rightarrow\rho_{x}. \label{eq:cq-channel}%
\end{equation}
Such a channel is known as a classical-quantum or cq channel for short. Two
parameters that characterize the performance of this channel for transmitting
classical data are the fidelity $F\left(  W\right)  \equiv\left\Vert
\sqrt{\rho_{0}}\sqrt{\rho_{1}}\right\Vert _{1}^{2}$ and the symmetric Holevo
information $I\left(  W\right)  \equiv H\left(  \left(  \rho_{0}+\rho
_{1}\right)  /2\right)  -\left[  H\left(  \rho_{0}\right)  +H\left(  \rho
_{1}\right)  \right]  /2$, where $H\left(  \sigma\right)  \equiv-$Tr$\left\{
\sigma\log_{2}\sigma\right\}  $ is the von Neumann entropy. A channel is near
perfect for transmitting classical data if $I\left(  W\right)  \approx1$ and
$F\left(  W\right)  \approx0$, and it is nearly useless for this task if
$I\left(  W\right)  \approx0$ and $F\left(  W\right)  \approx1$
(Ref.~\cite{WG11}\ gives a precise relationship between these channel parameters).

When encoding classical information for the channel $W$, we consider $N=2^{n}$
uses of it, so that the resulting channel is of the form: $x^{N}=x_{1}\cdots
x_{N}\rightarrow\rho_{x^{N}}\equiv\rho_{x_{1}}\otimes\cdots\otimes\rho_{x_{N}%
}, $ where $x^{N}$ is the classical input sequence and $\rho_{x^{N}}$ is the
output quantum state. Arikan's idea of channel combining and splitting extends
to this case by considering his encoding matrix $G_{N}$ \cite{A09}\ acting on
an input sequence $u^{N}$: $u^{N}\rightarrow\rho_{u^{N}G_{N}}, $ where
$u^{N}G_{N}$ is the binary vector resulting from multiplication of the row
vector $u^{N}$ by Arikan's encoding matrix $G_{N}$. We can then define the
\textquotedblleft split\textquotedblright\ channels $W_{N}^{\left(  i\right)
}$\ from the above \textquotedblleft combined\textquotedblright\ channels as follows:%
\begin{align}
W_{N}^{\left(  i\right)  }  &  :u_{i}\rightarrow\rho_{\left(  i\right)
,u_{i}}^{U_{1}^{i-1}B^{N}},\\
\rho_{\left(  i\right)  ,u_{i}}^{U_{1}^{i-1}B^{N}}  &  \equiv\sum_{u_{1}%
^{i-1}}\frac{1}{2^{i-1}}\left\vert u_{1}^{i-1}\right\rangle \left\langle
u_{1}^{i-1}\right\vert ^{U_{1}^{i-1}}\otimes\overline{\rho}_{u_{1}^{i}}%
^{B^{N}},\\
\overline{\rho}_{u_{1}^{i}}^{B^{N}}  &  \equiv\sum_{u_{i+1}^{N}}\frac
{1}{2^{N-i}}\rho_{u^{N}G_{N}}^{B^{N}}. \label{eq:averaged-cond-states}%
\end{align}
The interpretation of this channel is that it is the one \textquotedblleft
seen\textquotedblright\ by the bit $u_{i}$ if all of the previous bits
$u_{1}^{i-1}$ are available and if we consider all the future bits
$u_{i+1}^{N}$ as randomized. This motivates the development of a quantum
successive cancellation decoder \cite{WG11}\ that attempts to distinguish
$u_{i}=0$ from $u_{i}=1$ by adaptively exploiting the results of previous
measurements and quantum hypothesis tests for each bit decision.

The above synthesized channels polarize, in the sense that some become near
perfect for classical data transmission while the others become near useless.
This result follows in the quantum case by a martingale argument similar to
Arikan's:\ model the channel splitting and combining as a random birth
process, and it is possible to prove that the parameters $I(W_{N}^{\left(
i\right)  })$ and $F(W_{N}^{\left(  i\right)  })$ converge almost surely to
zero-one valued random variables in the limit of many recursions of the
encoding. The following theorem characterizes the rate with which the channel
polarization effect takes hold \cite{AT09,WG11}:

\begin{theorem}
\label{thm:fraction-good}Given a binary input cq channel $W$ and any
$\beta<1/2$, it holds that $\lim_{n\rightarrow\infty}\Pr_{I}\{\sqrt
{F(W_{2^{n}}^{\left(  I\right)  })}<2^{-2^{n\beta}}\}=I\left(  W\right)  $,
where $n$ indicates the level of recursion for the encoding, $W_{2^{n}%
}^{\left(  I\right)  }$ is a random variable characterizing the $I^{\text{th}%
}$ split channel, and $F(W_{2^{n}}^{\left(  I\right)  })$ is the fidelity of
that channel.
\end{theorem}

Suppose now that we can determine which channels are the good ones and which
ones are bad. In this case, we can construct a polar coding scheme by dividing
the synthesized channels according to the following polar coding rule:%
\begin{equation}
\mathcal{G}_{N}\left(  W,\beta\right)  \equiv\big\{i\in\left[  N\right]
:\sqrt{F(W_{N}^{\left(  i\right)  })}<2^{-N^{\beta}}\big\},
\label{eq:polar-coding-rule}%
\end{equation}
and $\mathcal{B}_{N}\left(  W,\beta\right)  \equiv\left[  N\right]
\setminus\mathcal{G}_{N}\left(  W,\beta\right)  $, so that $\mathcal{G}%
_{N}\left(  W,\beta\right)  $ is the set of \textquotedblleft
good\textquotedblright\ channels and $\mathcal{B}_{N}\left(  W,\beta\right)  $
is the set of \textquotedblleft bad\textquotedblright\ channels. The sender
then transmits the information bits through the good channels and
\textquotedblleft frozen\textquotedblright\ bits through the bad ones. A
helpful assumption for error analysis is that the frozen bits are chosen
uniformly at random such that the sender and receiver both have access to
these frozen bits. Ref.~\cite{WG11} provided an explicit construction of a
quantum successive cancellation decoder that has an error probability equal to
$o(2^{-\left(  1/2\right)  N^{\beta}})$---let $\{\Lambda_{u_{\mathcal{A}}%
}^{\left(  u_{\mathcal{A}^{c}}\right)  }\}$ denote the corresponding decoding
positive operator-valued measure (POVM), with $u_{\mathcal{A}}$ the
information bits and $u_{\mathcal{A}^{c}}$ the frozen bits.

\section{Polar codes for quantum communication}

\label{sec:polar-review-quantum} Wilde and Renes recently showed how to
construct quantum polar codes for sending quantum data over an arbitrary qubit-input
quantum channel
\cite{WR12} (before this, Renes \textit{et al}.~addressed the case of Pauli
channels \cite{RDR11}). The polar coding scheme in
Ref.~\cite{WR12} follows the general approach of Renes and Boileau
\cite{RB08}, in which they showed how to build up quantum codes from classical
codes for two different cq channels induced by the quantum channel of
interest. We briefly overview the approach given in Ref.~\cite{WR12}.

Suppose that a quantum channel $\mathcal{N}$ connects a sender to a receiver.
Consider that the sender can induce a cq \textquotedblleft
amplitude-basis\textquotedblright\ channel $W_{A}$ from $\mathcal{N}$ by
inputting a computational basis state $\left\vert z\right\rangle $\ depending
on some classical bit $z$:%
\begin{equation}
W_{A}:z\rightarrow\mathcal{N}\left(  \left\vert z\right\rangle \left\langle
z\right\vert \right)  . \label{eq:amp-cq}%
\end{equation}
Since the induced channel $W_{A}$\ is of the form in (\ref{eq:cq-channel}), it
is possible to construct a polar code for it of rate $I\left(  W_{A}\right)
$, consisting of the classical encoding $G_{N}$\ and a corresponding quantum
successive cancellation decoder as described in the previous section. Now
suppose that Alice shares a maximally entangled state $\left\vert
\Phi\right\rangle $ with Bob, where
\[
\left\vert \Phi\right\rangle \equiv2^{-1/2}\sum_{z\in\left\{  0,1\right\}
}\left\vert z\right\rangle \left\vert z\right\rangle .
\]
Alice can modulate her share of this state with $Z^{x}$ depending on some bit
$x$, so that the global state for Alice and Bob becomes $2^{-1/2}\left(
\left\vert 0\right\rangle \left\vert 0\right\rangle +\left(  -1\right)
^{x}\left\vert 1\right\rangle \left\vert 1\right\rangle \right)  $. If Alice
sends her share of the state through the channel, this process induces a cq
\textquotedblleft phase-basis\textquotedblright\ channel $W_{P}$\ of the
following form:%
\begin{equation}
W_{P}:x\rightarrow\left(  \mathcal{N}\otimes I\right)  \left(  Z^{x}\left\vert
\Phi\right\rangle \left\langle \Phi\right\vert Z^{x}\right)  .
\label{eq:phase-channel}%
\end{equation}
It is also possible to construct a polar code for $W_{P}$ of rate $I\left(
W_{P}\right)  $, again consisting of the classical encoding $G_{N}$ and a
quantum successive cancellation decoder.

The insight of Refs.~\cite{RDR11,WR12} is to build up a quantum polar code
from the polar codes for the above two channels. The sender and receiver can
compute offline which channel inputs will be good or bad for the amplitude and
phase channels. The sender then inputs the following types of qubits into the
different encoder inputs:

\begin{enumerate}
\item Information qubits into the synthesized channels that are good in both
amplitude and phase.

\item Ancilla qubits in the state $\left\vert 0\right\rangle $ into the
synthesized channels that are bad in amplitude and good in phase.

\item Ancilla qubits in the state $\left\vert +\right\rangle \equiv
2^{-1/2}\left(  \left\vert 0\right\rangle +\left\vert 1\right\rangle \right)
$ into the synthesized channels that are good in amplitude and bad in phase.

\item Shares of ebits in the state $\left\vert \Phi\right\rangle $ into the
synthesized channels that are bad in both amplitude and phase.
\end{enumerate}

The intuition behind this approach is that the receiver should be able to
recover quantum data if it is possible to recover classical data encoded into
complementary variables. Thus, the sender puts information qubits into the
synthesized channels which are good for both bases. The receiver will need
help in the decoding process in the form of frozen ancilla qubits for the
synthesized channels which are bad in one of the bases (this is the same as in
Arikan's classical polar coding scheme~\cite{A09}, with the exception that the
ancillas should be frozen in the basis that is bad). Finally, if both bases
are bad, a shared ebit is in some sense frozen in both bases, because Bob can
always predict the outcome of a Pauli $X$ or $Z$ measurement that Alice
performs on her end if she tells him which measurement she performs.

The net rate of this quantum polar coding scheme (rate of quantum
communication minus the rate of entanglement consumption) is equal to the
symmetric coherent information rate \cite{WR12,RDR11}:
\begin{align*}
I\left(  W_{A}\right)  +I\left(  W_{P}\right)  -1  &  =I\left(  A\rangle
B\right)  _{\mathcal{N}\left(  \Phi\right)  }\\
&  \equiv H\left(  B\right)  _{\mathcal{N}\left(  \Phi\right)  }-H\left(
AB\right)  _{\mathcal{N}\left(  \Phi\right)  }.
\end{align*}
The polar code operates as follows (see Ref.~\cite{WR12}\ for a detailed
discussion of the operation). Alice sends qubits as described above into a
coherent version of Arikan's encoder (this just consists of quantum
CNOT\ gates). Bob then exploits a coherent version of the quantum successive
cancellation decoder for the amplitude cq channel in order to coherently
decode the information qubits, placing them in a register $C^{N}$. Bob sends
the coherently decoded information qubits through the inverse of the coherent
polar encoder. This process induces the phase channel in
(\ref{eq:phase-channel}) from the point of view of the phase basis of the
qubits at the input of the encoder. Bob then acts with a coherent version of
the quantum successive cancellation decoder for the phase channels, coherently
placing the outcomes in a register $D^{N}$. Finally, he performs a linear
number of CNOT\ gates from the $C^{N}$ registers to the $D^{N}$ registers. The
result is that he will have decoded the information qubits with a fidelity
that is lower bounded by $1-2\sqrt{\epsilon_{A}}-2\sqrt{\epsilon_{P}}$, where
$\epsilon_{A}$ and $\epsilon_{P}$ are the error probabilities of the
respective polar codes for the cq amplitude and phase channels.

The polar encoder is efficient because it is the same network of CNOT gates in
Arikan's encoder (this part requires only $O\left(  N\log N\right)  $
operations). For general channels, it is still an open question to determine
if the polar decoder has an efficient implementation. For Pauli channels or
erasure channels, Renes \textit{et al}.~showed that the polar decoder has an
efficient implementation with $O\left(  N\log N\right)  $ operations
\cite{RDR11}, essentially because the induced cq amplitude and phase channels
are classical and they could thus exploit coherent versions of Arikan's
efficient polar decoder in the quantum polar decoder.

\section{Polar codes for private communication}

\label{sec:polar-review-private} An approach similar to the one above gives
polar codes for sending classical data privately over a quantum wiretap
channel \cite{WR12a}. Again, this approach follows the general approach of
Renes and Boileau \cite{RB08}\ for transmitting private classical data, in
which we build up such a protocol by considering two induced cq channels for
complementary variables. A quantum wiretap channel $\mathcal{N}^{A\rightarrow
BE}$\ has an input system $A$ for the sender, an output system $B$ for the
legitimate receiver, and an output system $E$ for the wiretapper. The goal in
this setting is for the sender to transmit classical data to the legitimate
receiver in such a way that the wiretapper obtains a negligible amount of
information about the input data.

We now review the private polar coding protocol from Ref.~\cite{WR12a}. The
first cq channel $\mathcal{M}_{A}$\ that we consider has Alice prepare a
quantum state $\rho_{z}$ depending upon a bit $z$ and feed this state into the
quantum wiretap channel $\mathcal{N}$:%
\begin{equation}
\mathcal{M}_{A}:z\rightarrow\mathcal{N}^{A\rightarrow B}(\rho_{z}).
\label{eq:cq-amp-wiretap}%
\end{equation}
By purifying the input state $\rho_{z}^{A}$ to $\left\vert \psi_{z}%
\right\rangle ^{AS_{1}}$ and the quantum wiretap channel $\mathcal{N}%
^{A\rightarrow BE}$ to the isometric map $U_{\mathcal{N}}^{A\rightarrow
BES_{2}}$ (where $S_{1}$ and $S_{2}$ are \textquotedblleft
shield\textquotedblright\ systems not possessed by the wiretapper
\cite{HHHO05PRL,HHHO09}), we can see that the above channel arises by tracing
over the wiretapper's system $E$ and the shield systems $S_{1}S_{2}$:
\[
z\rightarrow\left\vert \psi_{z}\right\rangle ^{BES_{1}S_{2}}\equiv
U_{\mathcal{N}}^{A\rightarrow BES_{2}}\left\vert \psi_{z}\right\rangle
^{AS_{1}}.
\]
The channel in (\ref{eq:cq-amp-wiretap}) is a cq channel, and as such, we can
construct a polar code for it with rate $I\left(  \mathcal{M}_{A}\right)  $
along with an encoder and quantum successive cancellation decoder \cite{WG11}.

The other cq channel that we consider is in some sense a virtual channel
because we only make use of it in order to reason about the security of our
private polar coding protocol (we do not actually exploit a decoder for it in
the operation of our protocol). This channel is a phase channel with quantum
side information. Suppose that Alice possesses an entangled state of the
following form:
\[
\left\vert \varphi\right\rangle ^{CAS_{1}}\equiv2^{-1/2}\sum_{z}\left\vert
z\right\rangle ^{C}\left\vert \psi_{z}\right\rangle ^{AS_{1}}.
\]
Alice could then modulate the $C$ system of the above state by applying the
phase operator $Z^{x}$ to it, depending on some input bit $x$. Suppose then
that she is able to transmit the $A$ system through the channel $\mathcal{N}%
^{A\rightarrow B}$ to Bob and the $C$ and $S_{1}$ systems through an identity
channel. The resulting cq phase channel to Bob is as follows:%
\begin{equation}
\mathcal{M}_{P}:x\rightarrow\mathcal{N}^{A\rightarrow B}((Z^{x})^{C}\left\vert
\varphi\right\rangle \left\langle \varphi\right\vert ^{CAS_{1}}(Z^{x})^{C}).
\label{eq:private-phase-channel}%
\end{equation}
Although we stated that this is a virtual channel, it is possible in principle
to construct a polar code for it with rate $I(\mathcal{M}_{P})$ along with a
quantum successive cancellation decoder \cite{WG11}.

The usefulness of the phase channels for privacy may not be readily apparent,
but an uncertainty relation from Ref.~\cite{RB08} clarifies the link. Indeed,
consider the following channel to the wiretapper:%
\begin{equation}
\mathcal{M}_{E}:z\rightarrow\mathcal{N}^{A\rightarrow E}(\rho_{z}).
\end{equation}
The uncertainty relation (Lemma 2 of Ref.~\cite{RB08}) for our case is as
follows:%
\begin{equation}
I\left(  \mathcal{M}_{P}\right)  +I\left(  \mathcal{M}_{E}\right)  =1.
\label{eq:security-uncertainty-relation}%
\end{equation}
The implication of this uncertainty relation is that if the phase channel
$\mathcal{M}_{P}$ to Bob is nearly perfect, then the amplitude channel
$\mathcal{M}_{E}$ is nearly useless to the wiretapper and vice versa.

We can then exploit the above uncertainty relation to construct a polar code
for private communication. The structure is similar to that in the previous
section, though the resources used are in some sense classical versions of the
quantum resources. The sender and receiver compute offline which channel
inputs will be good or bad for the amplitude or phase channels in
(\ref{eq:cq-amp-wiretap}) and (\ref{eq:private-phase-channel}), respectively.
The sender then inputs the following types of qubits into the different
encoder inputs:

\begin{enumerate}
\item Information bits to be kept private into the synthesized channels that
are good in both amplitude and phase.

\item Ancilla bits initialized to $0$ into the synthesized channels that are
bad in amplitude and good in phase.

\item Ancilla bits that are randomized ($0$ or $1$ with probability $1/2$)
into the synthesized channels that are good in amplitude and bad in phase.

\item Shares of secret key bits ($0$ or $1$ with probability $1/2$ and known
to both Alice and Bob) into the synthesized channels that are bad in both
amplitude and phase.
\end{enumerate}

The security of this approach follows from the uncertainty relation in
(\ref{eq:security-uncertainty-relation}). If Alice places the information bits
into the synthesized channels that are good in amplitude and phase, then Bob
will be able to recover them reliably since the channels are good in amplitude
and Eve learns only a negligible amount of information about them due to the
uncertainty relation in (\ref{eq:security-uncertainty-relation}) and the fact
that the channels are good in phase for Bob. The sender places ancilla bits
set to 0 in the channels that are bad in amplitude and good in phase in order
to help Bob in decoding the information bits. She places randomized bits into
the channels that are good in amplitude and bad in phase in order randomize
Eve's knowledge of them. Finally, she places secret key bits into the channels
that are bad for both variables in order to help Bob decode and to randomize
Eve's knowledge of the information bits.

The net rate of this private polar coding scheme (rate of private classical
communication minus the rate of secret key consumption) is equal to the
symmetric private information rate \cite{WR12a,RB08}:
\[
I\left(  \mathcal{M}_{A}\right)  +I\left(  \mathcal{M}_{P}\right)  -1=I\left(
\mathcal{M}_{A}\right)  -I\left(  \mathcal{M}_{E}\right)  .
\]
We can estimate the reliability and security of this scheme by considering the
performance of the constituent cq polar codes.

\section{Rigorous Bounds on Performance for the Quantum Erasure Channel}

\label{sec:erasure-bounds}

We now arrive at our first result, where we
provide rigorous bounds on the performance of quantum polar codes on
the quantum erasure channel. Our development here is both similar to and
extends that in Section~V-D~of Arikan \cite{A09}. Recall that a quantum
erasure channel takes a qubit in the state $\rho$ as input and outputs this
state with probability $1-\epsilon$ or an orthogonal erasure flag $\left\vert
e\right\rangle $ with the complementary probability:%
\begin{equation}
\rho\rightarrow\left(  1-\epsilon\right)  \rho+\epsilon\left\vert
e\right\rangle \left\langle e\right\vert . \label{eq:quantum-erasure-channel}%
\end{equation}

First, consider that in any given \textquotedblleft run\textquotedblright\ of
a quantum polar code of the form reviewed in
Section~\ref{sec:polar-review-quantum}, the block error probability $P_{e}$ is
as follows:%
\begin{equation}
P_{e}\equiv\Pr\left\{  \alpha_{\text{err}}\right\}  +\Pr\left\{
\phi_{\text{err}}\right\}  -\Pr\left\{  \alpha_{\text{err}},\phi_{\text{err}%
}\right\}  , \label{eq:quantum-block-err-prob}
\end{equation}
where $\alpha_{\text{err}}$ and $\phi_{\text{err}}$ are the events that an
error occurs during the decoding of the induced amplitude and phase channel,
respectively. (The above formula ensures that we count a quantum block error
when either an amplitude or phase error occurs, but that we do not overcount
when both errors occur.) By exploiting the law of total probability (that
$\Pr\left\{  \phi_{\text{err}}\right\}  =\Pr\left\{  \alpha_{\text{err}}%
,\phi_{\text{err}}\right\}  +\Pr\left\{  \overline{\alpha_{\text{err}}}%
,\phi_{\text{err}}\right\}  $, where $\overline{\alpha_{\text{err}}}$ is the
event that there is no amplitude decoding error), we have that%
\begin{align*}
P_{e}  &  =\Pr\left\{  \alpha_{\text{err}}\right\}  +\Pr\left\{
\overline{\alpha_{\text{err}}},\phi_{\text{err}}\right\} \\
&  =\Pr\left\{  \alpha_{\text{err}}\right\}  +\Pr\left\{  \phi_{\text{err}%
}\ |\ \overline{\alpha_{\text{err}}}\right\}  \Pr\left\{  \overline
{\alpha_{\text{err}}}\right\} \\
&  =\Pr\left\{  \alpha_{\text{err}}\right\}  +\Pr\left\{  \phi_{\text{err}%
}\ |\ \overline{\alpha_{\text{err}}}\right\}  \left[  1-\Pr\left\{
\alpha_{\text{err}}\right\}  \right] \\
&  =\Pr\left\{  \alpha_{\text{err}}\right\}  +\Pr\left\{  \phi_{\text{err}%
}\ |\ \overline{\alpha_{\text{err}}}\right\}  -\Pr\left\{  \phi_{\text{err}%
}\ |\ \overline{\alpha_{\text{err}}}\right\}  \Pr\left\{  \alpha_{\text{err}%
}\right\}  .
\end{align*}

When considering performance for the quantum erasure channel, there is a
symmetry that is helpful in simplifying the above expression. Consider that
both the induced amplitude and phase channels for the quantum erasure channel
in (\ref{eq:quantum-erasure-channel}) are classical erasure channels with
erasure probability $\epsilon$. Let $W_{N}^{\left(  i\right)  }$ be the
$i^{\text{th}}$ synthesized channel from the classical erasure channel, $\eta$
a threshold parameter between 0 and 1, and $\mathcal{A}\left(  \eta\right)  $
the set of information bits chosen by Arikan's polar coding
rule:\ $\mathcal{A}\left(  \eta\right)  \equiv\{i:Z(W_{N}^{\left(  i\right)
})\leq\eta\}$, with $Z$ the Bhattacharya parameter. Recall that the order of
these indices $i$\ is preserved for the induced amplitude channels but
reversed for the induced phase channels. If we assume that $\eta$ is the same
for both the induced amplitude and phase channels, then it follows that%
\[
\Pr\left\{  \alpha_{\text{err}}\right\}  =\Pr\left\{  \phi_{\text{err}%
}\ |\ \overline{\alpha_{\text{err}}}\right\}  .
\]
This assumption is reasonable because both of the induced channels are erasure
channels with the same erasure probability. By exploiting the above symmetry,
we have that%
\[
P_{e}=\Pr\left\{  \alpha_{\text{err}}\right\}  \left(  2-\Pr\left\{
\alpha_{\text{err}}\right\}  \right)  .
\]

Thus, in order to bound the block error probability $P_e$ from both above and below,
we only need to bound $\Pr\left\{  \alpha_{\text{err}}\right\}  $ from above
and below. In order to do so, we can exploit the following upper bound from
Section~V-B of Arikan \cite{A09}:
\[
\Pr\left\{  \alpha_{\text{err}}\right\}  \leq U\left(  \eta\right)  \equiv
\sum_{i\in\mathcal{A}\left(  \eta\right)  }Z(W_{N}^{\left(  i\right)  }).
\]
Then by using Arikan's recursive equations for the erasure channel in (38) of
Ref.~\cite{A09}, we can easily compute this upper bound on $\Pr\left\{
\alpha_{\text{err}}\right\}  $.

For a rigorous lower bound on $\Pr\left\{  \alpha_{\text{err}}\right\}  $, we
appeal to arguments similar to those in Section~V of Ref.~\cite{WG11}, since
one can always embed classical systems into quantum systems (also, recall that
Arikan only provided a heuristic lower bound in Section~V-D~of Ref.~\cite{A09}%
). We can write the error probability $\Pr\left\{  \alpha_{\text{err}%
}\right\}  $ as follows:%
\begin{multline}
\frac{1}{2^{N}}\sum_{u^{N}}\bigg(1-\text{Tr}\bigg\{\Pi_{\left(  N\right)
,u_{1}^{N-1}u_{N}}^{B^{N}}\cdots\Pi_{\left(  i\right)  ,u_{1}^{i-1}u_{i}%
}^{B^{N}}\cdots\Pi_{\left(  1\right)  ,u_{1}}^{B^{N}}\label{eq:error-term-q}\\
\rho_{u^{N}}\ \Pi_{\left(  1\right)  ,u_{1}}^{B^{N}}\cdots\Pi_{\left(
i\right)  ,u_{1}^{i-1}u_{i}}^{B^{N}}\cdots\Pi_{\left(  N\right)  ,u_{1}%
^{N-1}u_{N}}^{B^{N}}\bigg\}\bigg),
\end{multline}
where $\rho_{u^{N}}$ is the encoded state and $\Pi_{\left(  1\right)  ,u_{1}%
}^{B^{N}}$, \ldots, $\Pi_{\left(  N\right)  ,u_{1}^{N-1}u_{N}}^{B^{N}}$ are
projectors corresponding to quantum hypothesis tests to decode each bit
$u_{i}$. In the classical case, $\rho_{u^{N}}$ is just a distribution given by
the encoding and channel and $\Pi_{\left(  1\right)  ,u_{1}}^{B^{N}}$, \ldots,
$\Pi_{\left(  N\right)  ,u_{1}^{N-1}u_{N}}^{B^{N}}$ are indicator functions
corresponding to likelihood ratio tests. Consider the following operator
inequalities which hold for commuting operators $P_{1}$ and $P_{2}$ such that
$0\leq P_{1},P_{2}\leq I$:%
\begin{align*}
I-P_{1}P_{2}P_{1}  &  \geq I-P_{1},\\
I-P_{1}P_{2}P_{1}  &  \geq I-P_{2}.
\end{align*}
We can exploit these recursively in order to obtain the following lower bound
on the error term in (\ref{eq:error-term-q}), since the projectors
$\Pi_{\left(  1\right)  ,u_{1}}^{B^{N}}$, \ldots, $\Pi_{\left(  N\right)
,u_{1}^{N-1}u_{N}}^{B^{N}}$ all commute for a classical erasure channel:%
\begin{multline*}
\sum_{u_{1}^{i-1}}\frac{1}{2^{i-1}}\sum_{u_{i}}\frac{1}{2}\text{Tr}\left\{
\left(  I-\Pi_{\left(  i\right)  ,u_{1}^{i-1}u_{i}}^{B^{N}}\right)
\sum_{u_{i+1}^{N}}\frac{1}{2^{N-i}}\rho_{u^{N}}\right\} \\
=\sum_{u_{i}}\frac{1}{2}\text{Tr}\left\{  \left(  I-\Pi_{\left(  i\right)
,u_{i}}^{U_{1}^{i-1}B^{N}}\right)  \rho_{\left(  i\right)  ,u_{i}}%
^{U_{1}^{i-1}B^{N}}\right\}  \equiv P_{\text{err}}\left(  u_{i}\right)  ,
\end{multline*}
where in the last line we have exploited the states and notation defined in
Section~V of Ref.~\cite{WG11}. By recognizing that the quantity above is
equivalent to the error in discriminating the states $\rho_{\left(  i\right)
,0}^{U_{1}^{i-1}B^{N}}$ and $\rho_{\left(  i\right)  ,1}^{U_{1}^{i-1}B^{N}}$
in a quantum hypothesis test and recalling the relationship between hypothesis
testing error, the trace distance, and the fidelity (Bhattacharya parameter),
we have the following lower bound from Ref.~\cite{FG98}:%
\[
P_{\text{err}}\left(  u_{i}\right)  \geq\frac{1}{2}\left(  1-\sqrt
{1-Z(W_{N}^{\left(  i\right)  })^{2}}\right)  .
\]
Thus, since this bound holds for any index $i$, we obtain the following lower
bound on $\Pr\left\{  \alpha_{\text{err}}\right\}  $ for the quantum erasure
channel:%
\[
\Pr\left\{  \alpha_{\text{err}}\right\}  \geq L\left(  \eta\right)  \equiv
\max_{i\in\mathcal{A}\left(  \eta\right)  }\frac{1}{2}\left(  1-\sqrt
{1-Z(W_{N}^{\left(  i\right)  })^{2}}\right)  .
\]
This bound is also easy to compute by exploiting Arikan's recursive equations
for the erasure channel in (38)\ of Ref.~\cite{A09}.

The above development then leads us to the following theorem:

\begin{theorem}
\label{thm:bounds}The block error probability $P_{e}$\ for a quantum polar
code of the form reviewed in Section~\ref{sec:polar-review-quantum}\ when used
on a quantum erasure channel is bounded as follows:%
\[
L\left(  \eta\right)  \left(  2-U\left(  \eta\right)  \right)  \leq P_{e}\leq
U\left(  \eta\right)  \left(  2-L\left(  \eta\right)  \right)  .
\]

\end{theorem}

\begin{figure}[ptb]
\begin{center}
\includegraphics[
width=3.4411in
]{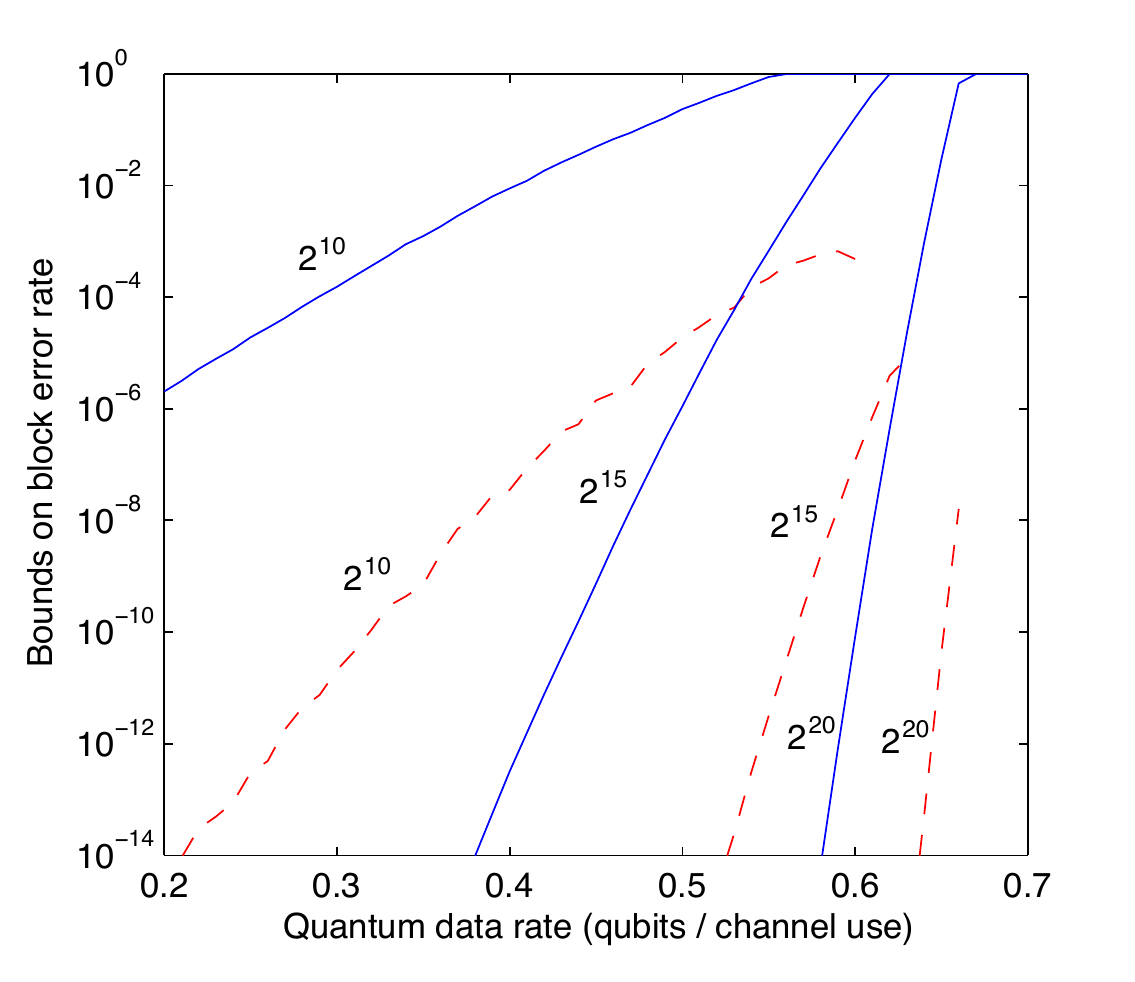}
\end{center}
\caption{The figure depicts the upper (blue solid curves) and lower (red
dashed curves) bounds from Theorem~\ref{thm:bounds}\ for a quantum erasure
channel with erasure probability $\epsilon=0.15$. These bounds determine the
trade-off between block error probability and rate for quantum polar codes
designed for a quantum erasure channel. The channel's quantum capacity
is equal to $1-2\epsilon=0.7$ qubits per channel use. Next to each curve, we
also show the block length for a quantum polar code to which the corresponding
bound applies.}%
\label{fig:erasure-bounds}%
\end{figure}

Figure~\ref{fig:erasure-bounds} depicts the bounds given by
Theorem~\ref{thm:bounds}. What we observe is that a large block length
$\approx2^{20}$ is required in order for a quantum polar code to operate in a
regime near the quantum capacity. Also, the margins between the upper and
lower bounds from Theorem~\ref{thm:bounds}\ are rather large. Thus, it is
clear that numerical simulations would be helpful in determining more accurate
estimates of the performance of quantum polar codes on the quantum erasure
channel.

\section{Numerical simulation results}

\label{sec:numerics}

We now describe the results of our numerical simulations of quantum polar
codes on quantum erasure, depolarizing, and BB84 channels. Our simulations rely
on two observations from Ref.~\cite{RDR11} (and noted in the previous section):

\begin{itemize}
\item The cq channels in (\ref{eq:amp-cq}) and (\ref{eq:phase-channel})
corresponding to the respective induced phase and amplitude channels result in
distinguishable states at the output (for erasure, depolarizing, and BB84 channels).

\item The performance of the resulting quantum polar code is directly related
to the performance of polar codes for the constituent amplitude and phase channels.
\end{itemize}

The first observation implies that the induced channels are effectively
classical. Specifically, for the quantum erasure channel with erasure
probability $\epsilon$, both the induced amplitude and phase channel
correspond to a classical erasure channel with erasure probability $\epsilon$.
The second observation implies that we can simulate the performance of a
quantum polar code for Pauli or erasure channels by simulating the performance
of Arikan's successive cancellation decoder for the constituent
channels. Importantly, this simulation is
efficient because Arikan's successive cancellation decoder
is efficient, requiring only $O (N \log N)$ operations.

Our Monte Carlo simulation proceeds by randomly generating information bits,
encoding them with the polar encoder, transmitting them over a given channel,
and then decoding with the successive cancellation decoder. In our simulations, we vary the choice of the
classical rates $R_{C}^{(\text{ph})}$ and $R_{C}^{(\text{amp})}$ for the
respective phase and amplitude channels. For the case of the quantum erasure
channel, it is optimal to choose these rates to be equal because the two
induced amplitude and phase channels are classical erasure channels with the
same erasure probability and hence have the same capacity.

\begin{figure}[ptb]
\centering
\includegraphics[width=\columnwidth]{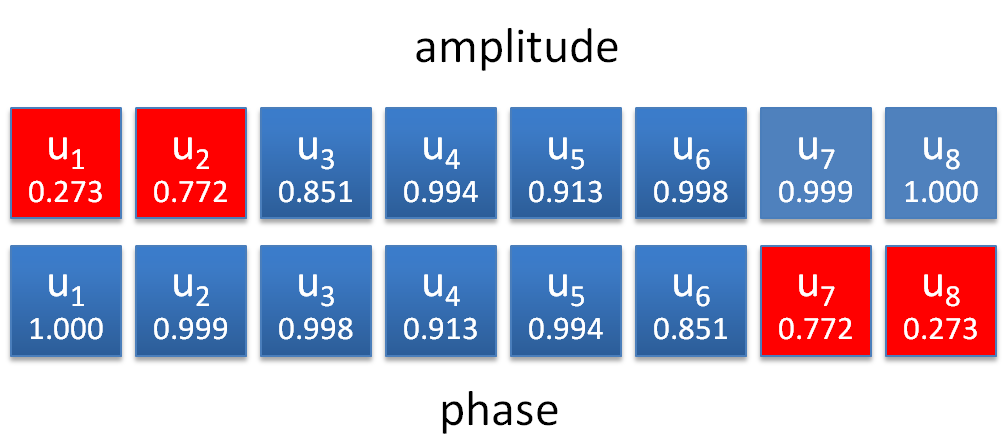}\caption{Diagram of the
information bits for the induced amplitude and phase channels for code length
$N=8$ and a classical rate $R_{C}= 0.75$. The numbers represent the symmetric
capacities for each virtual channel when $\epsilon=0.15$. The blue boxes
represent the `good' channels, which for our example are the virtual channels
with symmetric capacities above the lowest quartile. The quantum code then
chooses the four bits which are good for both amplitude and phase, resulting
in a rate $R_{Q}=0.5$ quantum polar code.}%
\label{fig:channels}
\vspace{-.15in}
\end{figure}

Figure~\ref{fig:channels} depicts the process by which we choose an example
quantum polar code with $N=8$ and classical rates  $R_{C}^{(\text{amp})}= R_{C}^{(\text{ph})}= 0.75$. The virtual channels' symmetric
capacities in the erasure case can be calculated explicitly and efficiently
via the recursive formulas in (38) of Ref.~\cite{A09}. The reason that the
symmetric capacities are reversed for the induced phase channel is that the
action of the quantum CNOT gates in the encoding are reversed when acting on
the phase basis (as first observed in \cite{RDR11}). We end up with four
channels ``good'' for both amplitude and phase and $R_{Q}=4/8=0.5$. In the
case where the bad virtual channels are completely clustered at the ends, it
holds that $R_{Q}=R_{C}^{(\text{ph})}+ R_{C}^{(\text{amp})}-1$. However, we
note that the symmetric capacities of the virtual channels are generally not
ordered sequentially ({\it e.g.}, $u_{5}$ has a larger symmetric capacity than
$u_{4}$) and in larger block lengths $N$, the good virtual channels become
distributed more non-contiguously, resulting in slightly higher quantum rates.

After randomly choosing the information bits $u_{1}^{N}$, the encoding is
performed by multiplying by the encoding matrix $G_{N}$ \cite{A09}, and the
codeword is transmitted through the channel according to its probability
transition matrix. For the amplitude channels, we then sequentially decode
$u_{1}$, $u_{2}$, $\cdots$.
We always decode any frozen bits correctly, and we decode the information bits to be either `0' or
`1' according to the likelihood ratio, which for each bit is calculated recursively in
$O(\log N)$ steps by incorporating the values of previously decoded bits
\cite{A09}. For the induced phase channels, we decode in the reverse order,
due to the reversal mentioned above.

We calculated the block error rates $\Pr\left\{  \alpha_{\text{err}}\right\}  $
and $\Pr\left\{  \phi_{\text{err}}\ |\ \overline{\alpha_{\text{err}}}\right\}  $
for the respective induced amplitude and phase channels
separately by performing $M$ such simulations of each, in each case comparing
with the actual encoded word. The block error rate for quantum data transmission is
equal to (\ref{eq:quantum-block-err-prob})
by counting an error whenever either the amplitude or phase decoding fails
(but not overcounting when both fail). For most cases, we ran $M=50,000$
simulations in order to get statistically significant results down to
$P_{e} \leq 10^{-4}$ (though for some larger $P_{e}$ cases we ran $M=5,000$, as this was more than sufficient to obtain small statisical error bars).

\subsection{Results for Quantum Erasure Channels}

\begin{figure}[ptb]
\centering
\includegraphics[width=\columnwidth]{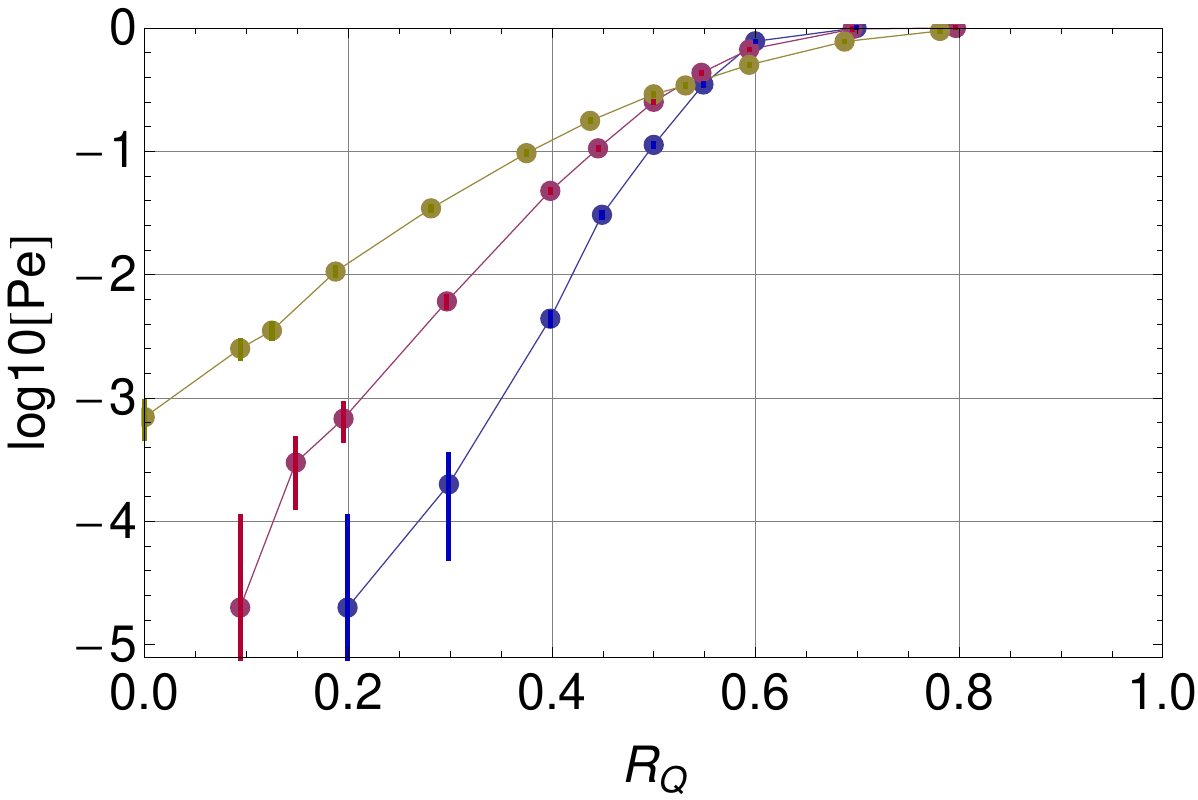}\caption{Block error rate
$P_{e}$ versus quantum data rate for a quantum erasure channel with
$\epsilon=0.15$ and different block lengths. Yellow, red and blue points show,
respectively, $N=64$, $256$, and $1024$. Points show simulated results and
lines are linear interpolations between points.
 The error bars here and in subsequent  figures reflect the 90\% confidence interval for the error rates given the number of simulations performed and errors observed.
The quantum capacity of this channel is equal to 0.7 qubits per channel use.}%
\label{fig:vsQrate}
\vspace{-.15in}
\end{figure}

\begin{figure}[ptb]
\centering
\includegraphics[width=\columnwidth]{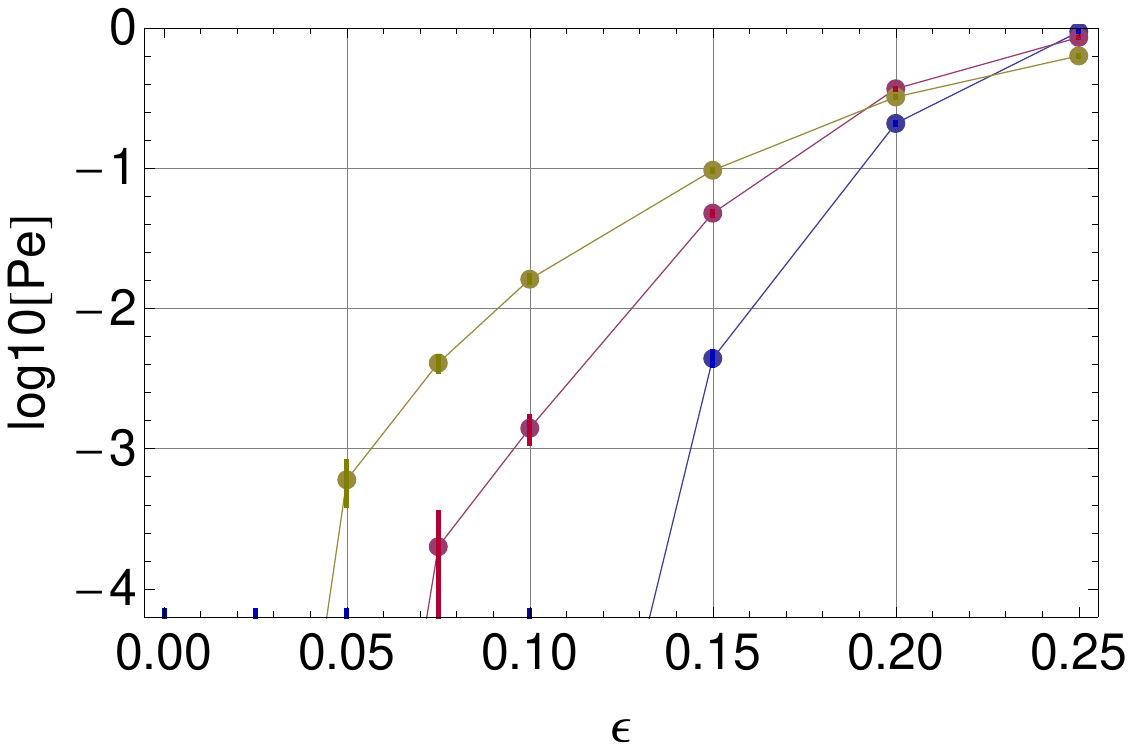}
\vspace{-.65in}
{\begin{flushleft}\bf\Large(a)\end{flushleft}}
\vspace{.2in}
\includegraphics[width=\columnwidth]{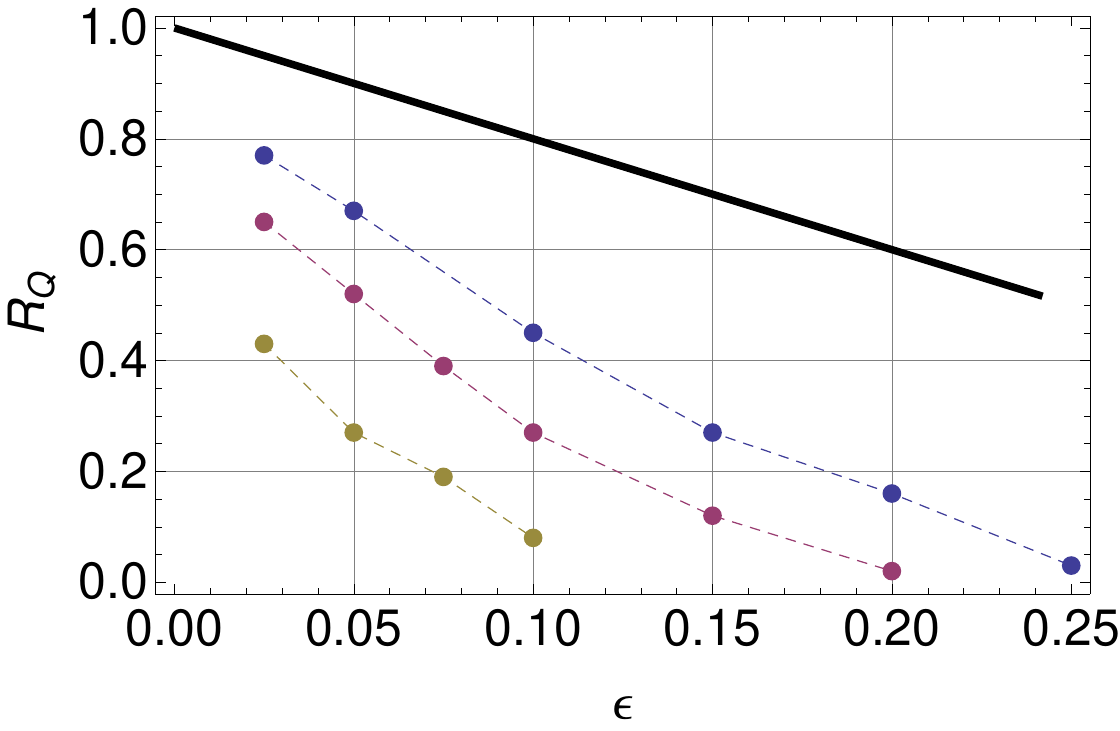}
\vspace{-.65in}
{\begin{flushleft}\bf\Large(b)\end{flushleft}}
\caption{Quantum erasure channel
simulation results. \textbf{(a)} Block error rate $P_{e}$ versus erasure
probability $\epsilon$ for $N=64$ (yellow), $N=256$ (red), and $N=1024$ (blue)
and quantum rate $R_{Q}=0.398$ (the $N=64$ points were for $R_{Q}=0.375$).
\textbf{(b)} Threshold quantum data rate $R_{Q}$ for block error rate
$P_{e} \leq 10^{-4}$ versus erasure probability $\epsilon$. The black curve shows
the quantum capacity $1-2 \epsilon$. }%
\label{fig:erasSims}
\vspace{-.15in}
\end{figure}

Figure~\ref{fig:vsQrate} depicts the block error rate versus quantum data rate
for three different code lengths $N=64$, $256$, and $1024$. All of the block
error rates start to bend down slightly below the quantum capacity for the
channel (which in this case is equal to 0.7 qubits per channel use)%
, and the longer block lengths
improve more rapidly at lower rates. Figure \ref{fig:erasSims}(a) then plots
the block error rate versus erasure probability $\epsilon$ for one particular
quantum data rate $R_{Q} \approx0.4$. Note that the exact rates for different
block lengths $N$ are slightly different due to a slightly increasing fraction of
overlapping phase-good and amplitude-good channels at larger block lengths.

Figure~\ref{fig:erasSims}(b) summarizes performance for a variety of erasure
probabilities $\epsilon$ and quantum data rates $R_{Q}$ by plotting the
threshold rate at which $P_{e} \leq 10^{-4}$. It also plots the quantum capacity
$1-2 \epsilon$ of the quantum erasure channel \cite{BDS97}. We observe that
for the selected block lengths, quantum polar codes can perform increasingly near
to capacity as the erasure probability $\epsilon$ decreases.

\subsection{Results for Quantum Depolarizing Channels}

We also simulated the performance of quantum polar codes for the quantum
depolarizing channel with depolarizing probability $p$, modeled as follows:
\[
\rho\to(1-p) \rho+ p/3\, (X \rho X + Y \rho Y + Z \rho Z) .
\]
In this case, the induced amplitude channel has orthogonal outputs, with
errors occuring only in the case of an $X$ or $Y$ flip. Thus, it is equivalent
to a classical binary symmetric channel (BSC) with flip probability $2p/3$.
The induced phase channel has four possible outputs (corresponding to the four
orthogonal Bell states), and it is thus equivalent to a classical channel with
the following probability transition matrix:
\[
\left(
\begin{array}
[c]{cccc}%
1-p & p/3 & p/3 & p/3\\
p/3 & p/3 & p/3 & 1-p
\end{array}
\right)  .
\]

One important subtlety for both the induced amplitude and phase channel for
the quantum depolarizing channel is that there is no explicit method for
calculating the symmetric capacities of each virtual channel
(as in the case of an erasure channel). From
Ref.~\cite{A09}, we know that there is a choice of good channels with
exponentially decreasing error probability for any rate below capacity, but
Ref.~\cite{A09} does not give an explicit prescription for choosing them. One
strategy from Ref.~\cite{A08}, which we employ here, is to calculate the symmetric
capacity $S$ of a channel and use an effective erasure probability $\epsilon=
1 - S$ to choose the good channels according the prescription for erasure
channels. It is not fully known how close to optimal this heuristic is.

Another important subtlety is that the induced amplitude and phase channels
have different symmetric capacities (with the phase channel's being slightly
larger). For this reason, it is better to choose the amplitude classical
rate slightly lower than the phase rate: $R_{C}^{(\text{amp})} < R_{C}^{(\text{ph})}$.  To account for this, we varied the rates $R_{C}^{(\text{amp})}$ and $R_{C}^{(\text{ph})}$ independently and chose the combination of these two rates which optimized $P_e$ for a given quantum data rate $R_Q$.   We found empirically that choosing $R_{C}^{(\text{amp})} = 0.82 R_{C}^{(\text{ph})}$ gave near optimal results across the channel parameters and rates we considered but varying around this choice gives slight improvements.  

\begin{figure}[ptb]
\centering
\vspace{-.02in}
\includegraphics[width=\columnwidth]{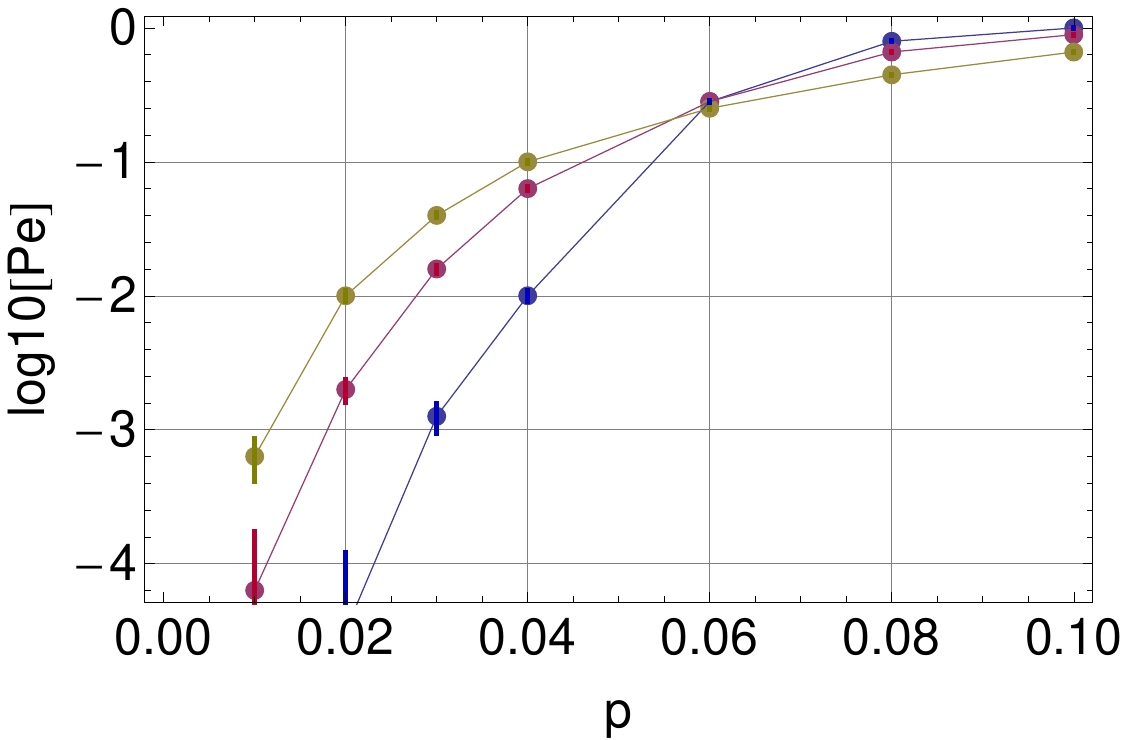}
\vspace{-.65in}
{\begin{flushleft}\bf\Large(a)\end{flushleft}}
\vspace{.1in}
\includegraphics[width=\columnwidth]{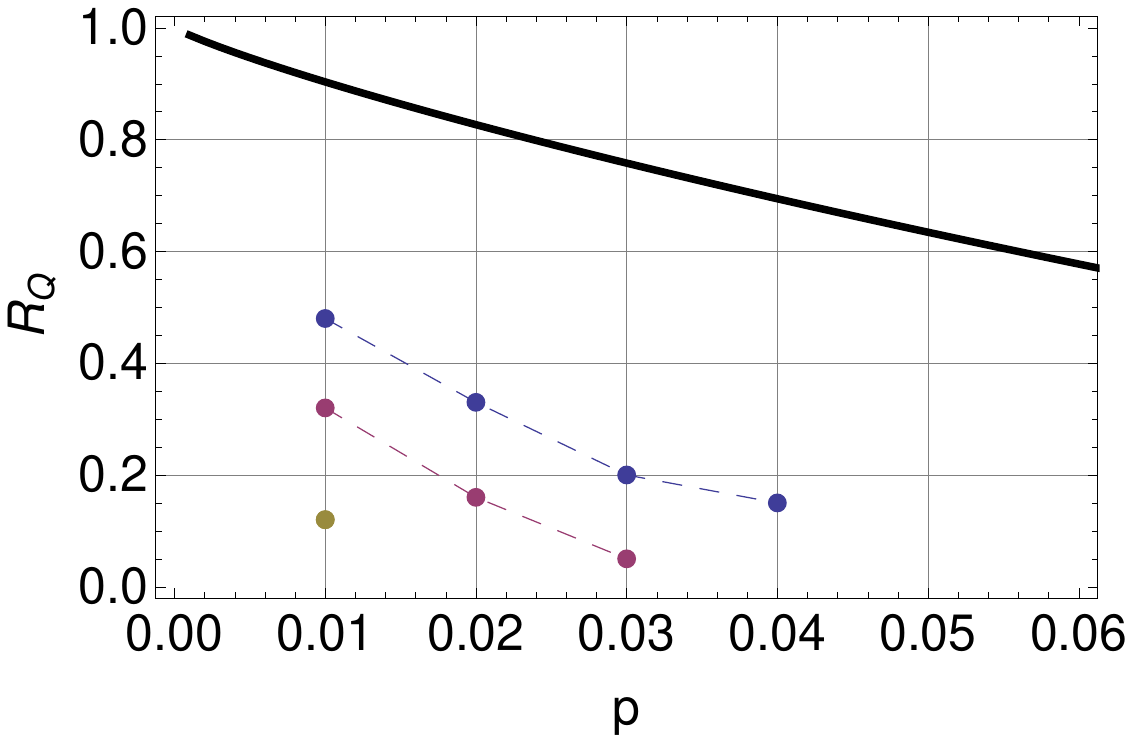}
\vspace{-.65in}
{\begin{flushleft}\bf\Large(b)\end{flushleft}}
\caption{Quantum depolarizing
channel simulation results. \textbf{(a)} Block error rate $P_{e}$ versus
physical error probability $p$ for $N=64$ (yellow), $N=256$ (red), and $N=1024$
(blue) and quantum data rate $R_{Q}=0.30$. \textbf{(b)} Threshold $R_{Q}$ for $P_{e} \leq 10^{-4}$ versus $p$.
The black curve shows the hashing limit $1-H_2(p)-p \log_{2}(3)$.}%
\label{fig:depolSims}
\vspace{-.15in}
\end{figure}

Figure~\ref{fig:depolSims}(a) plots the block
error rate $P_{e}$ for polar codes with quantum data rate $R_{Q} \approx0.30$, and
Figure~\ref{fig:depolSims}(b) plots the threshold for which $P_{e} \leq 10^{-4}$
along with the hashing bound $1-H_2(p)-p \log_{2}(3)$ (this is the rate
achievable by a random stabilizer code on the depolarizing channel).  We see the performance relative to the bounds is
not as strong as for the quantum erasure channel. Nevertheless, we obtain
fairly large quantum data rates with low block error rate $P_{e}$ and moderate block lengths
$N=1024$.

In Ref.~\cite{KHIS12}, Kasai {\it et al}.~presented results for quantum codes with code lengths in the range $N=4000$-$35,406$ and achieved quantum rates higher than ours at the $P_e<10^{-4}$ threshold.  We should perform simulations of higher $N$ to compare our coding scheme with theirs.  However, it appears from our evidence so far that at these code lengths, achievable quantum polar code rates will be lower than theirs, unless we exploit some improved method such as that in Ref.~\cite{TV10} for computing the good and bad virtual channels.

\subsection{Results for the BB84 Channel}

\begin{figure}[ptb]
\centering
\includegraphics[width=\columnwidth]{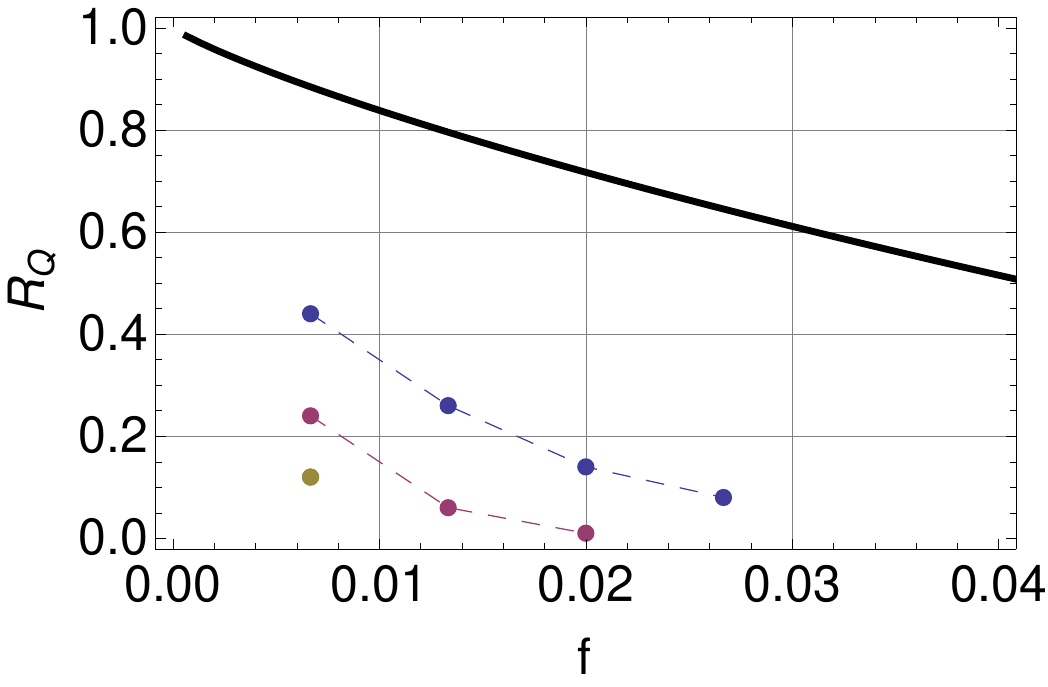}
\vspace{-.3in}\caption{BB84 channel simulation results.  Threshold $R_{Q}$ for $P_{e} \leq 10^{-4}$ versus flip probability $f$.
The black curve is the hashing limit $1-2 H_2(f)$.}%
\label{fig:bb84Sims}
\vspace{-.15in}
\end{figure}

Finally, we consider performance on the ``BB84 channel'' \cite{KR08}, which is a concatenation of a bit-flip and phase-flip channel, each having some equal flip probability $f$ (see Section 2.1.1 of Ref.~\cite{KR08} for the relevance of this channel in quantum key distribution). For the BB84 channel, the induced amplitude and phase channels are simply binary symmetric channels, each with error probability $f$. 
Because the two channels are equivalent, as in the erasure channel case, it is optimal to choose equal underlying rates:  $R_{C}^{(\text{amp})} =  R_{C}^{(\text{phase})}$.  Figure~\ref{fig:bb84Sims}  plots our results for quantum data rates achieving the $P_e<10^{-4}$ threshold.  We see quantum polar code performance is very similar to that for the depolarizing channel.

\section{Application to Private Communication}

\label{sec:private-app}

Our simulations above serve equally well as simulations of polar codes for
private classical data transmission over the channels we considered
(where here the eavesdropper gets access to
everything that the receiver does not obtain). This follows essentially from
Theorem~1 of Renes \cite{Renes08062011}, in which he proves that a lower bound
on Bob's guessing probability for the phase variable serves as an upper bound
for security against Eve with respect to the amplitude variable. Thus, the
parameter in (\ref{eq:quantum-block-err-prob}) serves as both a reliability and
security parameter for the protection of classical data sent through the channels we have considered.

\section{Discussion}

\label{sec:conclusion}

In summary, we have bounded the performance and conducted simulations of
quantum communication over the quantum erasure channel, by utilizing the
recently proposed quantum polar codes in Refs.~\cite{RDR11,WR12}. We found
high quantum data rates for moderate block lengths $N=1024$. We also
performed simulations of these codes over the depolarizing and BB84 channels.
In these cases, we found that quantum
polar codes for these block lengths still performed ably but gave performance
somewhat further from the hashing limits.

Going forward from here, it is important to explore the performance of larger block
lengths in order to compare the performance of quantum polar codes with other error correction
schemes. In the depolarizing and BB84 cases, it would be worthwhile to carry out an
analysis of the optimality of the erasure-matched code choice.
Also, one could exploit the techniques from
Ref.~\cite{TV10} in order to choose which virtual channels to send the information
bits through.

\bibliographystyle{IEEEtran}
\bibliography{Ref}

\end{document}